%% file: hipeac.tex

\documentclass[preprint]{sig-alternate}

\toappear{[Copyright notice will appear here once 'preprint' option is removed.]}

\usepackage{glossaries}
\usepackage{url}
\usepackage{color}

\input{variables}
\begin{document}

\input{titleAuthors}

\maketitle

\input{abstract}

\terms{capture and replay, iterative compilation}

\input{introduction}
\input{motivation}
\input{main}
\input{experiments.tex}
\input{relatedWork}
\input{conclusions}

\section{Acknowledgments}
This work was supported by the UK Engineering and Physical Sciences Research Council under grants
EP/H044752/1 (ALEA) and EP/M015793/1 (DIVIDEND).


\input{hipeac.bbl}
\end{document}

%% file: variables.tex
\newcommand\varGeneralIcSpeedup{40\%}
\newcommand\varGeneralIcSlowdown{70\%}
\newcommand\varIcBenchmarkName{fir filter}

\newcommand\varICCRSearchSequences{2000}

\newcommand\varCRIChighestSpeedup{57\%}
\newcommand\varCRICmostLowSpeedup{20\%}
\newcommand\varCRIClowestSpeedup{16\%}
\newcommand\varCRICsameEffectSequences{300}
\newcommand\varCRICaverageSlowdown{0.8\%}
\newcommand\varCRIChighestSlowdown{1.7\%}
\newcommand\varCRICvsIChighestNoise{less than 10$\mu$s}
\newcommand\varCRICvsICavgNoise{2.11$\mu$s}
\newcommand\varCRICavgReplaysInOneExec{10000}
\newcommand\varCRsizeFullMax{200MB}

\newcommand\varICtotalFlags{60}
\newcommand\varClangVersion{3.6.1}

\newcommand\EXTRAINFO1{}
\definecolor{lightgray}{gray}{0.6}

\newacronym{VMA}{VMA}{Virtual Memory Area}
\newacronym{COW}{CoW}{Copy On Write}

%% file: titleAuthors.tex

\makeatletter

\makeatother
\title{Iterative compilation on mobile devices
}
%
%
%
%
%

\numberofauthors{3} 
%
\author{
%
%
\alignauthor
Paschalis Mpeis\\
        \affaddr{University of Edinburgh}\\
		\email{p.mpeis@ed.ac.uk}
\alignauthor Pavlos Petoumenos\\
        \affaddr{University of Edinburgh}\\
        \email{ppetoume@inf.ed.ac.uk}
\alignauthor Hugh Leather\\
        \affaddr{University of Edinburgh}\\
        \email{hleather@inf.ed.ac.uk}
}

%% file: abstract.tex
\begin{abstract}

The abundance of poorly optimized mobile applications coupled with their increasing centrality in
our digital lives make a framework for mobile app optimization an imperative. While tuning
strategies for desktop and server applications have a long history, it is difficult to adapt them
for use on mobile devices. 

Reference inputs which trigger behavior similar to a mobile application's typical are hard to construct.
For many classes of applications the very concept of typical behavior is nonexistent, each user interacting
with the application in very different ways. In contexts like this, optimization strategies need to evaluate
their effectiveness against real user input, but doing so online runs the risk of user dissatisfaction when
suboptimal optimizations are evaluated. 

In this paper we present an iterative compiler which employs a novel capture and replay technique
in order to collect real user test cases and use it later to evaluate different transformations offline.
The proposed mechanism identifies and stores only the set of memory pages needed to replay the most
heavily used functions of the application. At idle and charging periods, this minimal state is combined with
different binaries of the application, each one build with different optimizations enabled.
Replaying the targeted functions allows us to evaluate the effectiveness of each set of optimizations
for the actual way the user interacts with the application.

For the BEEBS benchmark suite, our approach was able to improve performance of hot functions by up to
\varCRIChighestSpeedup{}, while keeping the slowdown experienced by the user on average at
\varCRICaverageSlowdown{}. By focusing only on heavily used functions, we are able to conserve
storage space by between two and three orders of magnitude compared to typical capture and replay
implementations.

\end{abstract}

%% file: introduction.tex
\vfill 
\section{Introduction}
The way we use computers has changed. Most of our interaction is not with powerful desktop or
laptop computers, but with mobile devices of limited capabilities. Despite the centrality of these devices in our
lives and the billions of users relying on them, we are still away from fully exploiting even the
limited processing resources they provide. As an example, the current Android compiler focuses on
compilation speed instead of high optimization. This means that all non-native Android applications
are barely optimized, wasting tremendous amounts of performance potential.

A technique that could be used for improving the user experience in such an enormous scale is iterative
compilation~\cite{coleman1995tile,aarts1997oceans}. It is a well established technique that can
readily outperform the standard optimization levels of a compiler~\cite{Fursin2005}. By intelligently 
constructing and testing different transformation sets, it quickly identifies sets with a
nearly optimal effect on performance.
Exhaustive offline approaches, successfully applied on embedded devices~\cite{Kisuki2000}, are
tailored for fixed applications, invariable input, and particular architectures. This is not the case
with mobile applications, as they are being regularly updated, operate on diverse architectures, and
their behavior is highly dependent on the way the users interact with them~\cite{Orso2003}. While
online approaches mostly solve the first two issues, they have to pay the price of evaluating poorly
performing transformations. Even self-adaptive algorithms~\cite{Park2011,Almagor2004,Agakova,Cavazos2007a}
will inevitably stumble upon transformations that can dramatically degrade a user's experience during
their early learning stages. And to make things worse, slight differences in the execution environment
from one execution to the next or interference from co-scheduled applications can drastically change
the apparent effectiveness of the evaluated transformations. Without means to execute again the exact
same test cases and use statistical methods to remove the measurement noise, we can easily end up selecting
suboptimal transformation sets.

To combine the different benefits of offline and online approaches, we propose the novel approach
of incorporating a lightweight capture and replay mechanism into our iterative compilation framework.
The capture and replay mechanism allows us to store the state of a running application just before the
execution of a heavily used function and use it anytime afterwards to replay that function. We can
then evaluate different binaries generated by the iterative compiler using our replay system instead
of letting the user interact directly with them.

Despite being an offline search, this process is personalized to the user as it is driven by
the captured test cases produced by the interaction of the user with the device. And because it is an
offline evaluation, poor performing transformations do not affect the user and we are able to handle
noise statistically. Evaluating transformations through replaying has the added benefit of only
executing the parts of the application we want to optimize, resulting in a much faster search of the
transformation space.

Finally, on the capture side, which is the only active component while the user interacts with
the device, we use novel ideas to minimize the overhead both in terms of runtime and storage so
that the user experience does not suffer. All of them translate into an iterative compilation
framework which searches fast, optimizes the application for the way the user uses it, and has
little to no overhead from the user perspective.

We evaluate our technique using real Android devices. The results indicate that our capture approach
is completely transparent to the user, with the potential slowdown introduced by our mechanism being
on average \varCRICaverageSlowdown{}. Using the captured state to drive iterative compilation results
in optimal transformation sets which outperform the highest optimization level of the compiler by
up to \varCRIChighestSpeedup{}. Finally, our results show that this performance speedup, obtained
inside our replay sandbox, translates into similar speedups when the application is executed normally.

The main contributions of this paper are: 
\begin{itemize}
  \item a novel and lightweight function capture mechanism, operating with little runtime or storage overhead
  \item an efficient, replay-based performance evaluation mechanism for different function versions of the same application
  \item a proof of concept offline iterative compilation system that performs personalized optimization, copes with architectural diversity, and gracefully handles the inherently noisy mobile environment
\end{itemize}

This paper is organized as follows.
The next section describes the motivation behind our work.
It is followed by Section~\ref{sec:mainCRIC}, which describes the design of our lightweight capture and offline iterative compilation through replaying.
Section~\ref{sec:experimentalSetup} describes the experimental setup, and is followed by an evaluation of our approach, at Section~\ref{sec:results}.
Related work can be found at Section~\ref{sec:relatedWork}.
Finally, our concluding remarks and future work are at Section~\ref{sec:conclusions}.

%% file: motivation.tex
\section{Motivation}\label{sec:motivation}

\begin{figure}[!htb]
  \centering
  \includegraphics[width=\linewidth]{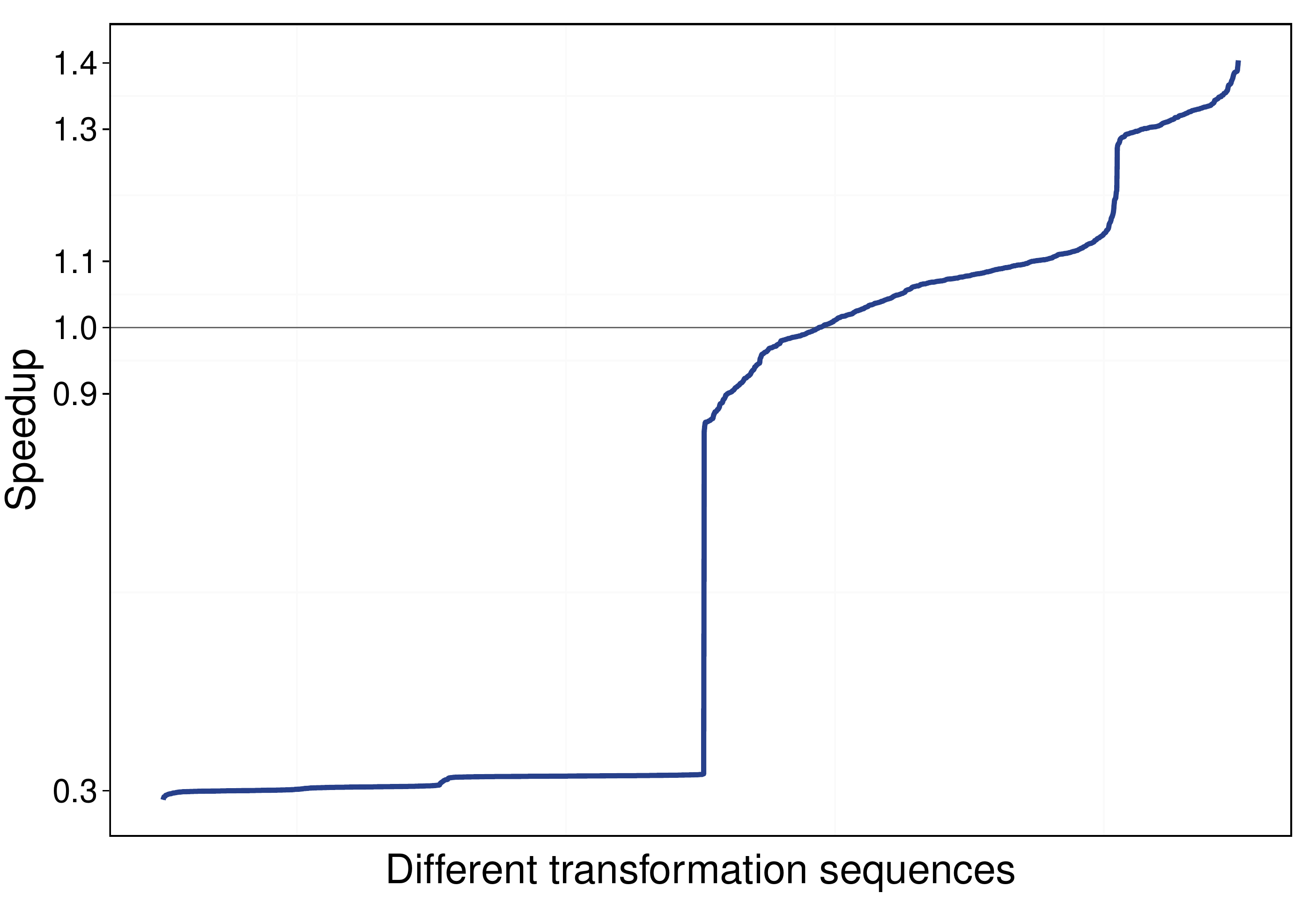}
  \caption
      {Speedup of \varICCRSearchSequences{} randomly picked transformation sets against Clang's
	  -O3 optimization level for the \textit{\varIcBenchmarkName{}} benchmark.}
   \label{fig:crDoubleEdged}
\end{figure}

Transformation sets tested by iterative compilation can outperform the highest optimization
level of a compiler, but can also dramatically degrade the performance. We see this double-edged
effect in Figure~\ref{fig:crDoubleEdged}, where \varICCRSearchSequences{} points randomly selected
from the massive transformation space were evaluated against the -O3 optimization level. While we
find many sets that outperform -O3 by up to \varGeneralIcSpeedup{}, we also stumble upon
many that unacceptably degrade the performance by up to \varGeneralIcSlowdown{}. Actually, for
more than half of the tested binaries, the mobile device ran unacceptably slow. This problem is
not specific to iterative compilation or the parameters used in our implementation. Peaking and
testing suboptimal transformation sets is an integral part of any learning process for
discovering well performing compiler optimizations. This means that any process which relies
on online evaluation of transformation sets cannot be used in the context of mobile devices. 

Using online instead of offline evaluation for an iterative compilation technique is bundled with
additional complications. Producing binaries for each distinct compiler optimization set is
a time consuming process. This can be done either online, directly affecting the user experience,
or in batch mode during periods of idleness, tying the speed of iterative compilation to the available storage capacity of the device. Both options are unsatisfactory.
And even if we managed to overcome this problem, the speed of iterative compilation would be limited
by the frequency of the user using the application. Dependencies like these would slow down the
evaluation time from minutes to months, even for the tiny fraction of the space that we visited for
the purposes of Figure~\ref{fig:crDoubleEdged}.
 
Our approach avoids all these problems by removing the need for online evaluation, while still
optimizing the application for real user test cases. A lightweight capture
of the state of the application during the invocation of heavily used functions allows us to
reproduce offline the way the user really interacts with the application. Then, distinct
transformation sets are used to produce different binaries, prepared in a way that pointers
to data or code in the captured state will remain valid across all the binaries. The captured
function invocation can then be directly re-executed to evaluate the transformation's effectiveness.

Fusing the replay mechanism with iterative compilation, allows us to push all undesired overheads
of the latter at periods where the user experience is not affected, e.g when the device is idle and being charged.
As a consequence, the performance noise is both minimal and manageable,
and the transformation evaluation is efficient since only the concerned code is being re-executed.
The architectural diversity is inherently addressed as the application is evaluated on the device
itself. Finally the conducted search is personalized to the user since their captured test cases are used to drive each replay. The next section describes in greater detail our approach.

%% file: main.tex
\section{Iterative compilation through capture and replay}\label{sec:mainCRIC}
This section describes the main components of our approach. Our capture and replay implementation operates completely at the Linux kernel userspace, and it is source language agnostic. The capture mechanism, described at Section~\ref{sec:capture}, is transparent to the users. It is designed in a way that minimizes in runtime and storage overhead. The collected information can be used by our replay mechanism, described at Section~\ref{sec:replay}, to re-execute at idle-and-charging periods the captured state of the application. Combined with appropriate link-time strategies this allows us to replay and evaluate different binaries corresponding to different compiler transformation sets. We seamlessly integrate it with a proof of concept iterative compiler, described at Section~\ref{sec:iterComp}, to showcase the manifold benefits of our approach.

\subsection{Lightweight function capture}\label{sec:capture}
Our capture targets \textit{hot functions}, whose execution frequencies and computational intensities
make them worth optimizing. It occurs on the first invocation of the selected hot function and its
overhead is negligible. Like existing approaches, it operates completely on the user space, requires
minimal instrumentation, is independent from an application's source language, and requires no kernel
or runtime system modifications.

Since the capture mechanism is the only part of our approach that is active while the user interacts
with the device, we need it to introduce as little runtime and storage overhead as possible. To achieve
that, we repurpose two kernel mechanisms, \gls{COW} and page-level protection. With
the former mechanism we have to copy the state of the application only when it's modified and even then
all copying is done efficiently in kernel space. With the latter mechanism we can identify the pages
actually used by the application and capture only them. Overall, we are able to significantly reduce
the size of the capture while by merging much of our functionality into existing kernel functionality
we minimize the amount of extra processing needed and the amount of kernel-userspace communication.

More specifically, right before a hot function's invocation, we parse through the \texttt{/proc} filesystem
the list of \glspl{VMA} used by the application and we call \texttt{fork} to duplicate the process.
The two processes, have separate address spaces that initially point to the same set of physical pages.
When the parent tries to modify a page, the efficient \gls{COW} mechanism will duplicate it keeping the
original contents in the child's copy of the page. This way the initial state that needs to be copied
for the capture is reduced to the bare minimum, i.e. the pages modified by the parent. As an added benefit,
the cost of this copying is offloaded to the kernel, further reducing the overhead of the capture. 
To completely remove competitiveness between the two processes, we reduce the priority of the child,
allow it to continue execution after the parent's completion, and put it in an idle state until further notice.

\begin{figure}[!htb]
  \centering
  \includegraphics[width=\linewidth]{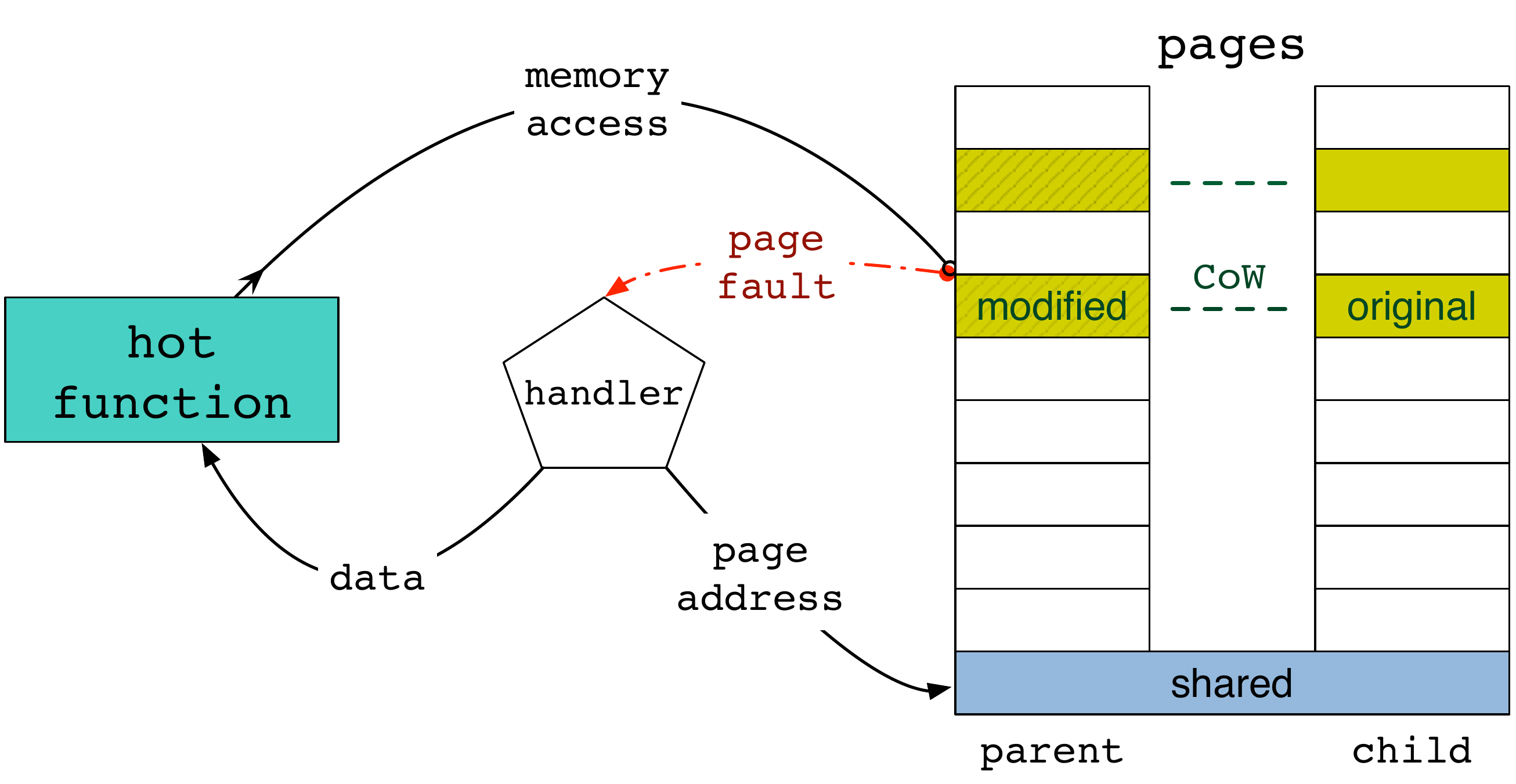}
  \caption
   {Identify the set of pages accessed by a hot function. The kernel's \gls{COW} mechanism efficiently preserves an original copy in the child process.}
   \label{fig:identifyPages}
\end{figure}

To pinpoint the set of pages that are used by a hot function, the parent process installs a fault handler and removes all the access rights from its pages.
Later on, when the parent tries to access a memory location, as shown in Figure~\ref{fig:identifyPages}, a page fault will be raised.
Our handler will process the fault by saving the address of the faulting page in a shared memory area. After that the access rights of the page are restored.
This mechanism guarantees that all accessed pages will be marked as such, while the overhead of doing so will be minimal.
For most applications, where a decent amount of spatial locality is expected, the space covered by the
accessed pages should be in the same order of magnitude as the space actually used by the application.
This keeps the amount of state that we save close to the minimum that needs to be saved. For the same
reason, the number of page faults and the associated overhead should be low.

The final step is to actually save the state on the device's storage. When the parent process is
about to complete the execution of the hot function, it signals the child process to proceed with
the capture. The child then simply reads the list of accessed pages from the shared memory and
stores into permanent storage the equivalent pages residing in its memory space.

\subsection{Replaying multiple versions of functions}\label{sec:replay}
The replaying mechanism itself is rather straightforward. The application binary and the saved state
are read into memory and a jump is performed just before the invocation of the hot function. The
function is re-executed until completion, then performance statistics are saved and the process
terminates.

\begin{figure}[!htb]
  \centering
  \includegraphics[width=\linewidth]{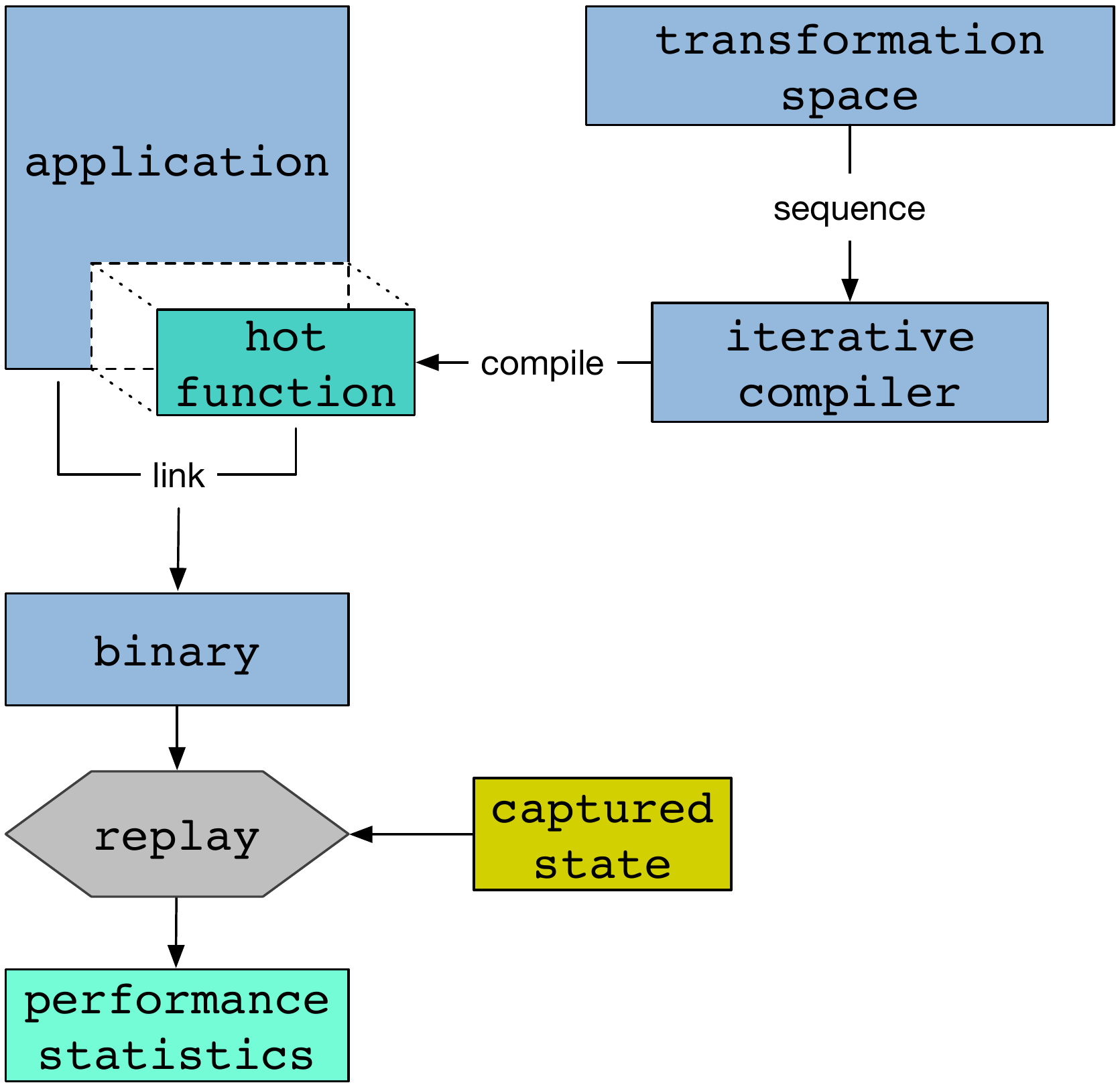}
  \caption
   {Applying offline iterative compilation through replaying. The hot function is repeatedly recompiled for evaluating new transformation sets, and then linked to a binary. 
Then, the captured data are used by the binary to replay the hot function, before finally storing the performance statistics.}
   \label{fig:iterCompiler}
\end{figure}

What is unique in our approach is the need to replay different binaries than the one used
during the capture, so that we can evaluate how different compiler transformation sets affect performance. 
To achieve that we tinker with the way that our binaries are built.
We see in Figure~\ref{fig:iterCompiler}, that an object code of a hot function is repeatedly transformed and plugged in into an existing object.	
Producing a correct executable is not straightforward. One issue is that
transformations might replace particular generic calls to faster architectural specific ones. This
causes changes in the \texttt{Procedure Linkage Table} and \texttt{Global Offset Table} sections of
the executable and cascading changes in the following sections. The result is that global variables
and function pointers now have different addresses than the ones in the version used for the capture.
In other words, when we restore the saved state, we will fail to load all this information in
the positions expected by the new binary. Another similar issue is that many transformations affect
the length of the hot function. Again, this causes cascading changes in the object code of following
functions, making function pointers in the saved state invalid.

To address the first problem, we used a helper object file at time of linking. It consists of a function
that is never actually called and has in its body dummy calls to functions that can be potentially
introduced by transformations. However, it is marked as being called, which tricks the assembler to
include the entries of the function calls to the linkage tables. For the second problem, we have extended
the linker script, so that empty space is added immediately after the hot function and the next function
always starts from the same offset. Such tinkering causes each binary to be in a consistent state with
the stored state in spite having a part of its object code transformed.

\subsection{Replay-based iterative compilation of hot functions}\label{sec:iterComp}
A proof of concept iterative compiler was implemented to evaluate our approach. The task of the
iterative compiler is to build replay-capable binaries by applying transformations solely to the
code of particular hot functions, as depicted in Figure~\ref{fig:iterCompiler}. It works with
applications written in the C language. Hot functions are extracted from the rest of the source code
and subsequently translated by the underlying Clang~\cite{clang} compiler using randomly picked
transformation sets. The resulting object file is linked against the previously compiled object
files of the rest of the sources in the way described in Section~\ref{sec:replay}.

The compiler evaluates each produced binary through the replay mechanism. Since we use the input
state saved during the application's normal use, our search algorithm is personalized for each user. 
Additionally, conducting the search at idle periods alleviates the user from having to suffer the
effects of poorly performing transformations, while the performance noise is kept to a minimum. By
performing re-compilation and re-execution on the device itself, the architectural diversity is
inherently addressed, and our focus on hot functions allows the speedup of transformation evaluations.


%% file: experiments.tex
\section{Experimental setup}\label{sec:experimentalSetup}
We have evaluated our approach by running a series of experiments on an actual mobile device.
The device we used was a Motorola Nexus 6 2014, which ran Android version 5.1; the latest available version for the device as of the time of writing. The device has a Qualcomm Snapdragon 805 processor, which is powered by four 2.7GHz Krait 450 cores.
To minimize the performance noise, we kept all cores online and hardwired the processor's frequency to its highest value. 

We have used benchmarks found in the open source BEEBS benchmark suite~\cite{webpageBeebsBenchmarks}.
It contains applications written in C, focused on embedded systems. We preferred this suite
over others, as the source code of most benchmarks was in a simple format that allowed an easy integration with our prototype system. With a manual profiling phase, using the Callgrind~\cite{callgrind} profiling tool, the hot functions of an application were identified and manually extracted from the rest of the sources. Due to the differences in computation power between embedded systems and current mobile devices we have increased the input in some of the applications, in order to make the hot functions computationally intensive, and therefore optimize-worthy.

Our iterative compiler runs on the device itself and uses the Clang driver~\cite{clang}, version \varClangVersion{}. It searches the space of possible transformations
by randomly building transformation sets of arbitrary length, from a list of \varICtotalFlags{}
transformation flags.
For each flag, we flip a coin to decide whether to include it or not. If the answer is positive, we flip again to decide on the parameter of the flag. Most of the
flags accept a boolean parameter, however, some might be more complex, which can dramatically
increase the search space. Even if we ignore this, such a space consists of $2^{\varICtotalFlags{}}$ distinct
points. In our experiments, we have visited only a tiny fraction of this space, by constructing
\varICCRSearchSequences{} random transformation sets.
We have made available online\footnote{\label{footnote:expSources}Benchmark sources and flags: \url{http://git.io/v4dlD}}
the compiler's space, the benchmarks, and the inputs, for the reproducibility of the results.

Our iterative compiler uses statistically rigorous techniques to deduce to the best performing
transformations. Once we have replayed each binary version 10 times and collected its runtimes, we
perform outlier removal using the the robust \textit{median absolute deviation} method. With the
\textit{two-side student's t-test}, we compare the results between different transformation
sets. The sets that are found to be best can then be used to build an optimal executable,
by recompiling once more the hot functions of an application.

All other measurements were repeated 30 times. Where applicable, we calculated the 95\% confidence intervals, and where appropriate, we removed outliers.


\section{Results and Analysis}\label{sec:results}
To test our technique we performed four sets of experiments. The first set examines the performance
benefits produced by our iterative compiler and whether the replay sandbox affects the effectiveness
of the evaluated transformations. The second set explores the overheads introduced by the capture
mechanism and whether they result in noticeable degradation of the user experience. The third set
of experiments compares our approach against traditional capture-and-replay techniques in terms of
storage overhead. Finally, we examine how the search speed of iterative compilation is affected by
being able to replay only the targeted hot function.

\subsection{Optimizing applications with iterative compilation through replaying}
This work is not about improving but about enabling iterative compilation for a large class of
applications. Nevertheless, it is important to show that our approach is able to optimize these
applications and that near optimal sets of transformation discovered through replay are still
near optimal when the application is executed normally.

\begin{figure}[!htb]
  \centering
  \includegraphics[width=\linewidth]{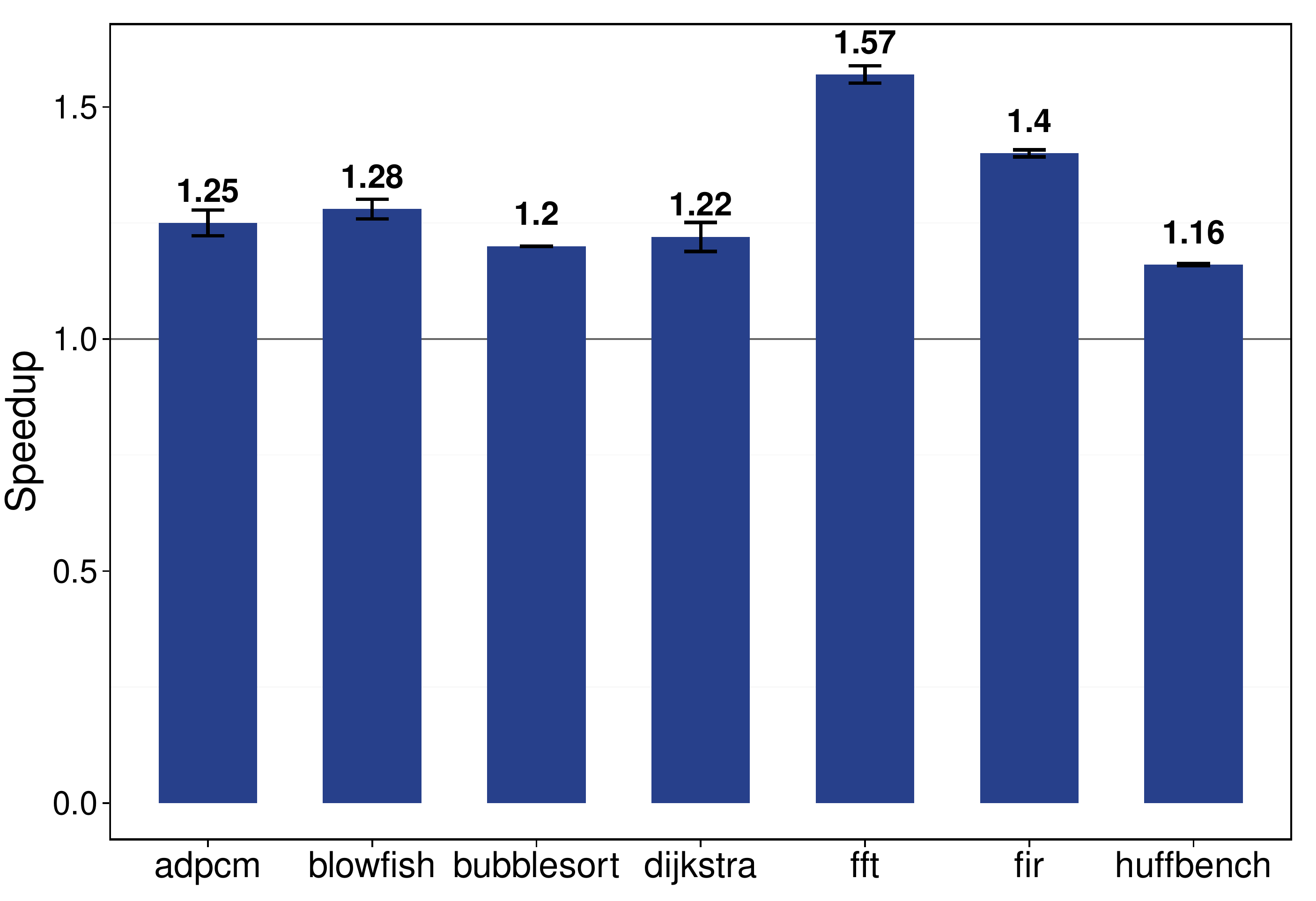}
  \caption
  {Speedup obtained by iterative compilation through replaying against the -O3 optimization level of Clang, after visiting \varICCRSearchSequences{} random points of the transformation space.}
   \label{fig:cricSpeedup}
\end{figure}

In Figure~\ref{fig:cricSpeedup}, we see the speedup for the best binary evaluated by our iterative
compilation system for the hot functions of seven benchmarks of our benchmark suite. All speedups are calculated
against the performance of the hot functions of a binary that was produced using the -O3 optimization level. Despite visiting
only a tiny fraction of the transformation space with a random search, the technique was still
able to increase the performance of all the benchmarks' hot functions, and for almost all by at least \varCRICmostLowSpeedup{},
compared to the highest optimization level of the compiler. The highest speedup obtained was
\varCRIChighestSpeedup{} for \texttt{fft}, while the lowest was \varCRIClowestSpeedup{} for
\texttt{huffbench}. A more extensive or intelligent search would produce even higher speedups,
but achieving worthwhile speedups with so little effort is a significant result on its own.

\begin{figure}[!htb]
  \centering
  \includegraphics[width=\linewidth]{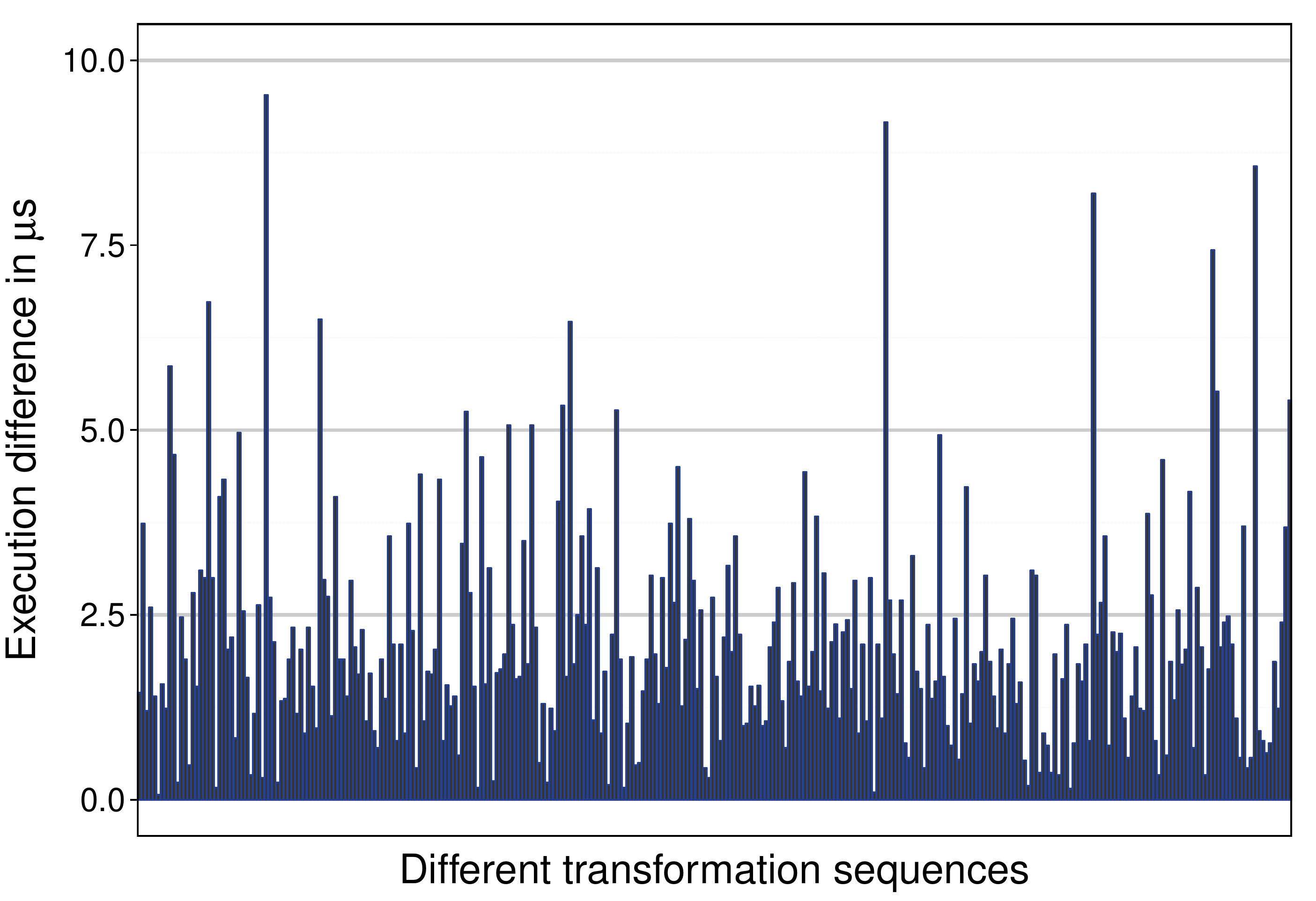}
  \caption
   {Difference in execution time of \varCRICsameEffectSequences{} transformations applied to the
	   hot function of the \varIcBenchmarkName{} benchmark between replayed and regular execution.}
   \label{fig:crNotAffectsIc}
\end{figure}

Figure~\ref{fig:crNotAffectsIc} depicts the differences in execution time for the hot function of
the \texttt{fir} benchmark when executed normally and when using the replay mechanism. For \varCRICsameEffectSequences{}
randomly picked transformations, the highest difference was \varCRICvsIChighestNoise{}, which is
within acceptable noise levels. The average difference in execution time was \varCRICvsICavgNoise{},
which is even smaller. It is clear that the link-time strategies described at Section~\ref{sec:replay},
that enable replaying of different versions of the hot function, do not interfere with the effectiveness
of the applied transformations. In other words, transformations found offline through replaying still
produce similar speedups when the application is executed normally.

\subsection{Transparent capturing}
While we can use replay and iterative compilation to eventually produce faster running executables,
it's still important that the whole process does not affect negatively the user experience, even
temporarily. To achieve that, we need to make sure that capture, the only part of our system running
while the user interacts with the device, introduces minimal overheads.

\begin{figure}[!htb]
  \centering
  \includegraphics[width=\linewidth]{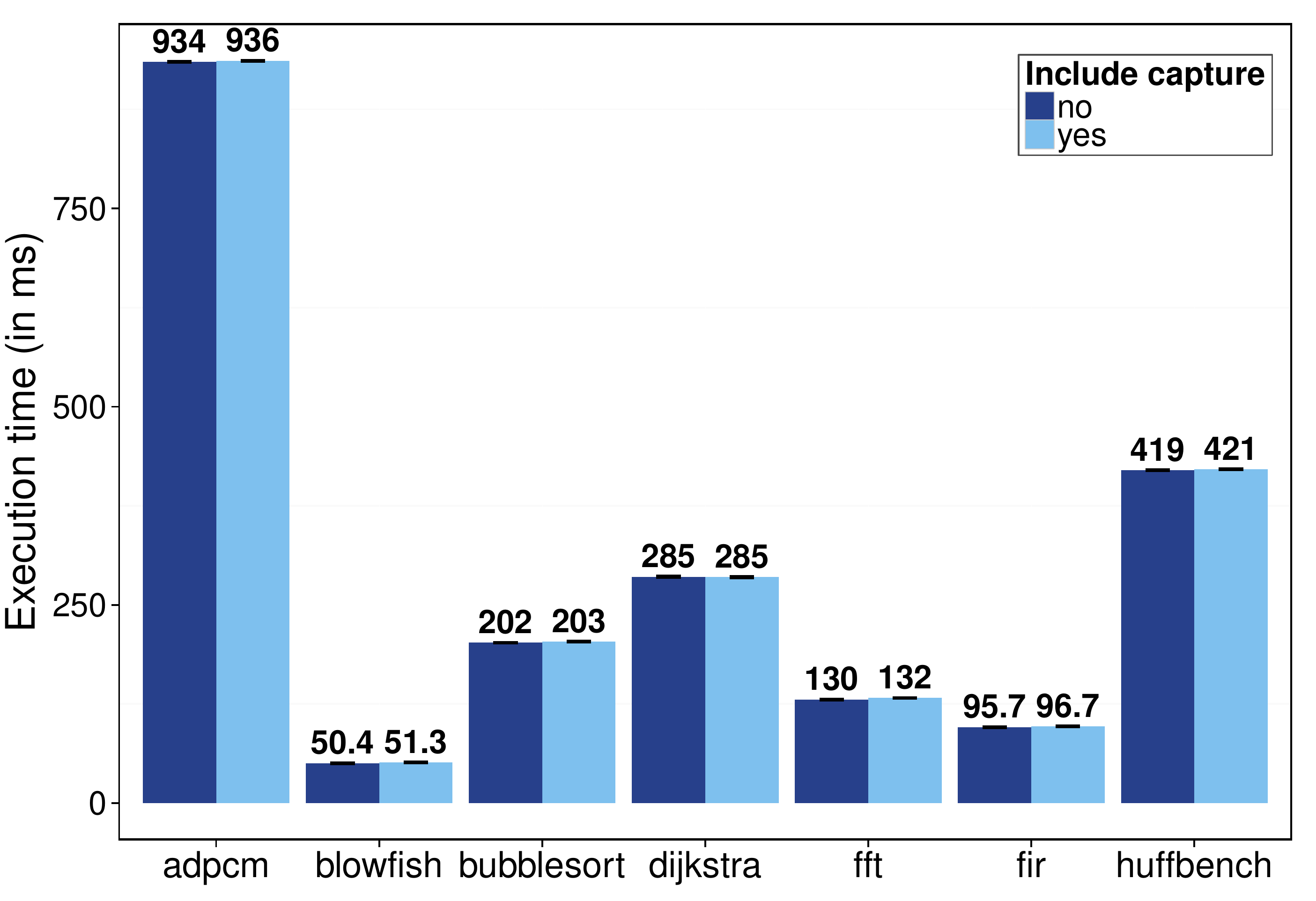}
  \caption
   {Regular execution compared against execution that were partially captured. The introduced slowdown does not surpass 2\% and is unlikely to affect the user experience.}
   \label{fig:crOverhead}
\end{figure}

In Figure~\ref{fig:crOverhead}, we see the runtime of each benchmark under normal and captured
execution. The slowdown introduced by the capture mechanisms is negligible compared to the execution
time of the tested benchmarks. The highest observed slowdown was \varCRIChighestSlowdown{} while the
average slowdown was just \varCRICaverageSlowdown{}.

Nearly all of our overhead comes from parsing the list of \glspl{VMA} and the call to \texttt{fork}
at the beginning of the capture process. The overhead of fault handling is negligible, since the
total number of faults is small even for memory-intensive applications.
In the worse case scenario, it can be as many as the pages owned by the application.
The \gls{COW} mechanism has a similarly small
overhead. \gls{COW} is used only when a page is modified and even then all memory management and
copying is done exclusively in kernel space, keeping the process fast. Overall, slowdowns that
small are unlikely to be noticed by the user, making our capture mechanism fit for mobile devices
used by real users. 

\subsection{Conserving space}
We saw in the previous subsection that parsing the list of \glspl{VMA} is responsible for a
significant part of the overall overhead of our capture and replay mechanism. The reason we need to
pay this cost is because parsing the \glspl{VMA} and removing the access rights from all the pages
are necessary for identifying the pages used by the hot function. With this information, we can
drastically cut down the size of the state that we need to capture and store.

\begin{figure}[!htb]
  \centering
  \includegraphics[width=\linewidth]{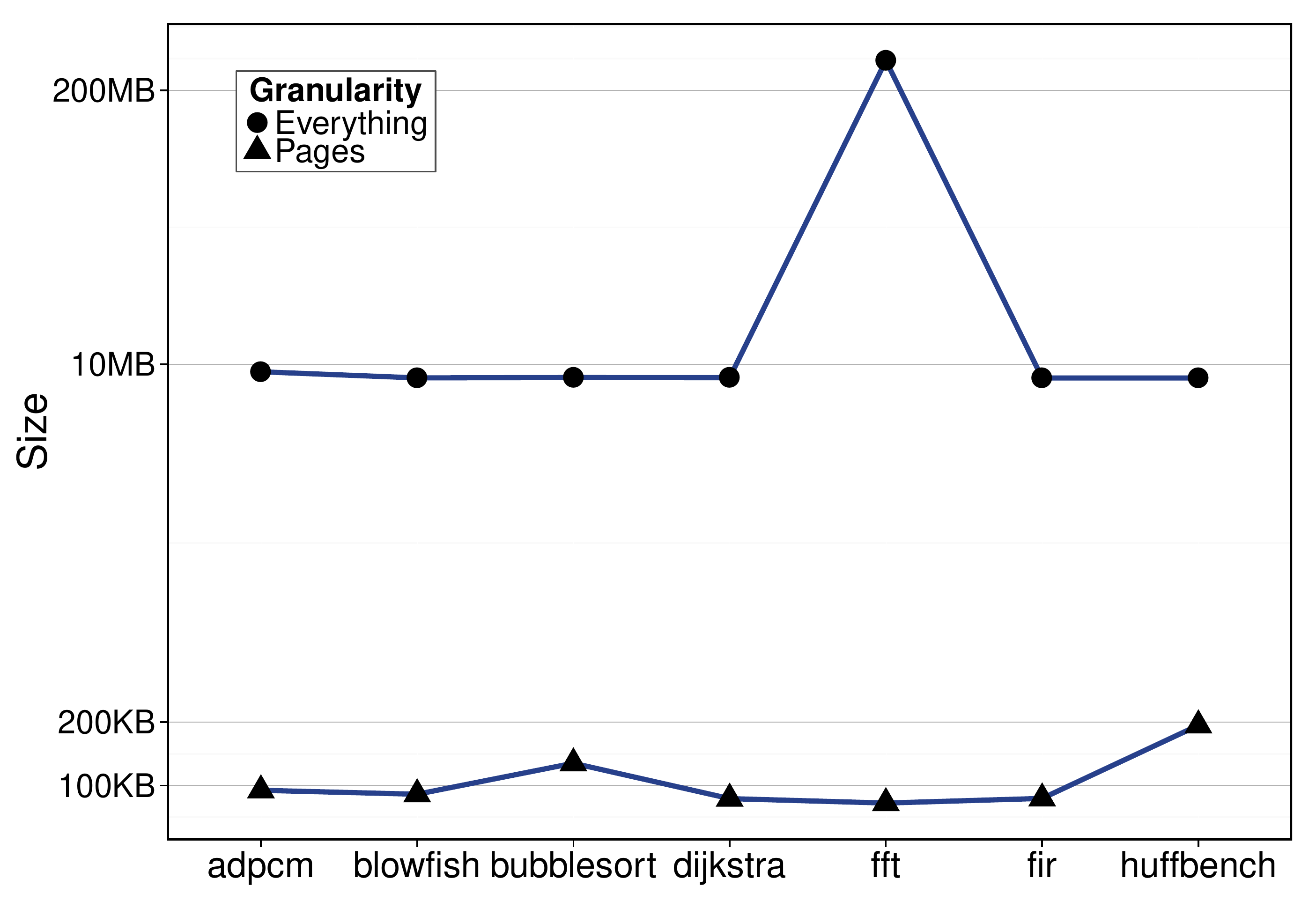}
  \caption
   {Comparing full captures against page-granularity captures, relevant to a hot function. There were at least two orders of magnitude storage space savings for all benchmarks.}
   \label{fig:crFullVsPages}
\end{figure}

In Figure~\ref{fig:crFullVsPages}, we see the space savings by capturing only the pages that are
used by the hot function, instead of saving all the state of the process. For all benchmarks, less
than 200KB of information were found to be necessary. For full captures that amount was at least
two orders of magnitude greater. An extreme example is the \texttt{fft} benchmark, where more than
\varCRsizeFullMax{} were stored during the full capture, with our space savings reaching three orders
of magnitude. Reducing the storage requirements of our approach, without imposing a performance
penalty, is particularly important. With all data easily fitting in the device's internal storage,
the available capacity of the device is hardly affected from the perspective of the user and there
is no need for storing potentially private data in external storage. 

\subsection{Speeding up the evaluation of transformations}
Using a replay mechanism for iterative compilation creates another potential benefit. In typical
iterative compilation techniques, each binary version is executed and evaluated as a whole, even
if only a small part of the program is targeted by iterative compilation. Using replay instead, we
directly jump at the point where the evaluation of the hot function starts and we terminate immediately
afterwards.

\begin{figure}[!htb]
  \centering
  \includegraphics[width=\linewidth]{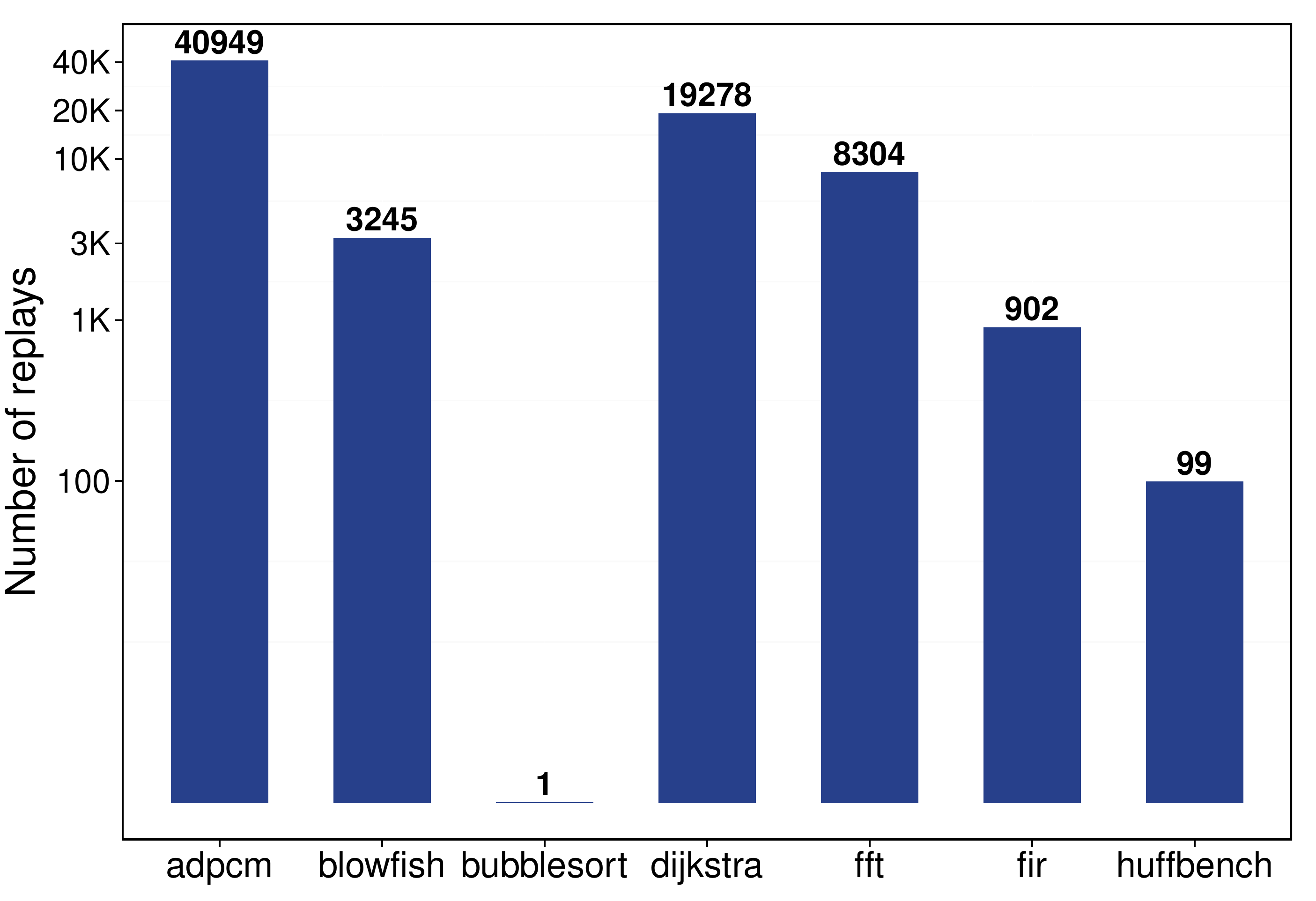}
  \caption
   {The number of times a hot function can be replayed, in the time that is required for a single full execution. Many replays can fit in that time slot for almost all of the benchmarks.}
   \label{fig:replaysInSingleExec}
\end{figure}

Figure~\ref{fig:replaysInSingleExec} shows how many replays can be performed in the runtime of a
single full execution of the benchmark or equivalently how much iterative compilation is accelerated
when using replay. Apart from bubblesort, where all processing is done by a single invocation of the
hot function, all other benchmarks can be sped up hundreds to tens of thousands of times. The average
number of replays fitting in a single execution of the benchmark are \varCRICavgReplaysInOneExec{}.
As far as iterative compilation is concerned, each set of replays will be a statistically sound
evaluation of a transformation. Having the ability to test more transformations during the same time
slot, can potentially lead to significant time and energy savings for the technique, as the only code
that is executed is the one that is being optimized.

%% file: relatedWork.tex

\section{Related work}\label{sec:relatedWork}
Iterative compilation is a well studied technique \cite{kisuki1999feasibility}, used for around two decades.
Early applications of the technique were successfully applied for particular embedded devices~\cite{Aarts1997, Bodin1998}.
An exhaustive offline search was performed once and then the discovered transformations were being used to optimize once and for all the device. The frequently updating mobile applications, their variable input, and the underlying architectural peculiarities render this approach impracticable on mobile devices.

Fursin \textit{et al}~\cite{Fursin2005a} presented a method that exploits online performance stability to evaluate multiple evaluations during the same run. Our approach offers a similar capability, without having to perform an online assessment of a transformation that can be detrimental to performance. Moreover, our approach has no dependency to the number of times a particular block of code is being executed during the normal execution.

Cooper \textit{et al}~\cite{Cooper1999} have used a genetic algorithm to search for transformations that would reduce the code size. The algorithm was able to evolve over time, by avoiding non beneficial transformations. Many other techniques based on machine learning have been proposed since then~\cite{Almagor2004, Agakova, Cavazos2007a, fursin2008milepost}, however, none of which can work transparently on mobile devices. The algorithms' learning curve is to blame, as it is their need to also learn poor transformations, transformations which when evaluated will degrade the user's experience.
Collective mind~\cite{fursin2014collective} is a framework that provides an abstraction layer for auto-tuning.
It tackles the chaos between different hardware, operating systems, software tools, search algorithms, and compilers versions, by allowing researchers to crowdsource information and cross-validate their findings.

Capture and replay is also a well studied technique. Approaches with similar emphasis on minimizing
the amount of captured state have used instrumentation~\cite{Orso2005, Xu2007, Jha2013} for logging
and later replaying application events, such as memory accesses. This allows them to store state at
granularity of a variable, which reduces the amount of stored state to the minimum possible. On the
other hand, this approach incurs significant overheads as all memory accesses are intercepted by
the logging mechanism. Another disadvantage of such an approach is the dependency to the source
language due to instrumentation. jRapture~\cite{Steven2000} has avoided instrumentation, by
modifying the Java API. While this is more transparent to the application, it still has the
disadvantages of the previous approaches.
CRIU~\cite{criu} is a capture and replay implementation that exploits the Linux kernel capabilities to avoid instrumentation or dependencies on the source language.
Its drawback, from our perspective, is that it is designed for live process migration in data centers. This means that full state captures are required, an extremely wasteful tactic in the context of mobile devices.

%% file: conclusions.tex
\section{Conclusions and Future work}\label{sec:conclusions}
In this paper we investigated whether iterative compilation can be applied for optimizing applications in the context of mobile devices.
Typical iterative compilation techniques when used online are associated with prohibitively high overheads, intolerable by the users of such devices.
Even algorithms that progressively reduce that overhead will slowdown the application significantly during their early learning stages.
Still, online iterative compilation offers the opportunity to evaluate transformations on real user test cases, which is hard to accurately emulate
offline.

Our solution for using the benefits of online iterative compilation without the drawbacks is to
combine iterative compilation with a novel capture and replay mechanism. This mechanism is designed with
the limitations of mobile devices in mind: limited processing power, limited storage, low latency
interaction between the user and the device. Specifically, we target only heavily used functions of
the application and we only capture the state needed for replaying them. This way, most of the state
of the application is ignored, reducing the needed storage by at least two orders of magnitude. To
identify the state needed for replaying and to copy it efficiently, we take advantage of existing
kernel mechanisms, keeping the performance overhead of our approach low. A custom replay mechanism
can then re-execute such hot functions, even in the presence of code transformations.
We integrate the
capture and replay mechanism with an iterative compiler. It takes the application state captured
online, while the user was interacting with the application, and uses it to replay and evaluate 
alternative binary version of the application. Evaluations take place when the device is not in use and being charged, so they
have no negative effect on the user experience.

We showed that iterative compilation through replay is able to outperform the highest optimization
level of a compiler by up to \varCRIChighestSpeedup{}. Our replay mechanism does not alter the impact 
of compiler transformations, as the difference between the regular runtime and the runtime during
replay is on average only \varCRICvsICavgNoise{}. Moreover, our approach allows multiple transformation
evaluations in the time slot of a single full execution. The overhead introduced by the initial capture
is not noticeable to the users, being on average \varCRICaverageSlowdown{}. Comparing it with full
capture alternatives, we are able to conserve storage space by between 2 and 3 orders of magnitude.

The ability to perform personalized optimization is an aspect of our proposed mechanism that deserves more attention.
The way a user interacts with a device might change over time requiring us to recalibrate the selected set of transformations.
Multiple users might have similar usage patterns, which could be used to enhance the transformation search.
We are currently evolving our approach to work with user-interactive mobile applications, which would allow us to further explore this territory.